\documentclass{webofc}

\usepackage{caption}
\usepackage{listings}
\usepackage[draft]{minted}
\usepackage[caption=false]{subfig}
\usepackage{hyperref}
\usepackage{graphicx}
\usepackage{tikz}

\usepackage[varg]{txfonts}   
\newminted{cpp}{gobble=2, numbersep=3pt, escapeinside=||, xleftmargin=8mm, framesep=2mm, frame=leftline, baselinestretch=1, fontsize=\scriptsize, linenos=false}
\newmintinline{cpp}{}

\usetikzlibrary{shadows, arrows, fit, positioning, matrix, shapes}
\usetikzlibrary{decorations.markings}

\definecolor{my_yellow}{RGB}{255, 253, 217}
\definecolor{my_orange}{RGB}{255, 127, 0}
\definecolor{my_lightblue}{RGB}{105, 186, 249}
\definecolor{my_purple}{RGB}{150, 154, 219}
\definecolor{my_green}{RGB}{90, 194, 160}

\tikzset {
  bigbox/.style = {draw, thick, fill=gray!10, rounded corners, rectangle},
  box/.style = {draw, thick, minimum height=0.8cm, minimum width=1.5cm, rounded corners, rectangle, fill=white, anchor=south},
  model/.style = {draw, thick, fill=white, text centered, minimum height=3em, minimum width=4em, rounded corners, drop shadow},
  user/.style = {draw, thick, ellipse, fill=white, text centered, minimum height=3em, minimum width=5em, drop shadow},
  line/.style = {->, thick, color=black, shorten <=2pt, shorten >=2pt, >=stealth'},
  dashed/.style = {->, dash pattern=on 3pt off 3pt, color=gray, shorten <=2pt, shorten >=2pt, >=stealth'},
  plain/.style = {minimum width=1em},
  arcnode/.style 2 args={
    decoration={
                 raise=#1,             
                 markings,   
                 mark=at position 0.5 with {\node[inner sep=0] {#2};}
            },
            postaction={decorate}
    }
}

\pgfdeclarelayer{background}
\pgfdeclarelayer{foreground}
\pgfsetlayers{background,main,foreground}

\begin{document}
\title{Continuous Performance Benchmarking Framework for ROOT}
%
%

\author{\firstname{Oksana} \lastname{Shadura}\inst{1}\fnsep\thanks{\email{oksana.shadura@cern.ch}},
        \firstname{Vassil} \lastname{Vassilev}\inst{2}\fnsep\thanks{\email{vvasilev@cern.ch}},
        \firstname{Brian Paul} \lastname{Bockelman}\inst{1}\fnsep\thanks{\email{brian.bockelman@cern.ch}}
}

\institute{University of Nebraska Lincoln, 1400 R St, Lincoln, NE 68588, United States
\and
           Princeton University, Princeton, New Jersey 08544, United States
          }
\abstract{%
  
Foundational software libraries such as ROOT are under intense pressure to avoid software regression, including performance regressions. Continuous performance benchmarking, as a part of continuous integration and other code quality testing, is an industry best-practice to understand how the performance of a software product evolves. We present a framework, built from industry best practices and tools, to help to understand ROOT code performance and monitor the efficiency of the code for several processor architectures. It additionally allows historical performance measurements for ROOT I/O, vectorization and parallelization sub-systems.


}
\maketitle
\section{Introduction}

In the last decade, software development has grown in size and complexity. To counteract potential defects due to this complexity, industry best practices include extensive quality testing, often done at regular intervals (e.g., ``nightly builds") or before a change is accepted into the code base. Regular quality assurance processes for software development is widely referred to as \textit{continuous integration}. Continuous integration often relies on fixed test suites and focuses on finding incorrect behavior. Correct behavior may be as simple as a pre-determined result of a library function for a specific fixed input data. In some domains, however, non-scalable (or otherwise poorly-performing) correct behavior can be as much as a defect as incorrect behavior.

As the field of high-energy physics (HEP) must deal with the large CPU and I/O costs of processing and analyzing billions of events, software performance is a critical and, in particular, for core libraries such as ROOT \cite{root}. It is important to understand changes in performance over time given performance regressions may be classified as critical for experiments' software stacks. Recently, metrics such as thread scalability or usage of vectorization have become critical to the field.

While performance is fundamental for software toolchains, performance analysis remains a laborious job requiring expertise. Reasoning about performance is difficult as it may require multiple hardware platforms (often using expensive or hard-to-acquire hardware), running with multiple compilers, or experimenting with specific setups. Through the use of continuous integration, a significant fraction of the work required for a reasonable understanding of performance can be automated and presented in the form accessible to non-experts. In this paper, we discuss the challenges of developing a continuous \textit{performance} monitoring system for ROOT; argue for the need of small-scale benchmarks; and discuss ways to visualize the obtained data. We describe one such solution, the \textit{ROOTBench} system providing:
\begin{itemize}
\item an extensible system tracking different performance metrics for the ROOT framework;
\item speculative performance evaluation of proposed changes;
\item rich and customizable visualizations and aids for performance analysis.
\end{itemize}

This remainder of this paper discusses background and related work in continuous performance monitoring (Section \ref{sec:background}), Information Flow in ROOTBench (Section \ref{sec:info-flow}), and usage of the ROOTBench system (Section \ref{sec:usage}).  Finally, we discuss our current results and conclusions in Sections \ref{sec:results} and \ref{sec:conclusions}, respectively.
 
\section{Background and Related Work} \label{sec:background}

We divide our requirements for a continuous performance benchmarking framework into three categories: (1) performance monitoring primitives, (2) visualization features, and (3) data storage. The performance monitoring primitives should have at least facilities for measuring elapsed time, memory footprint and user-defined counters. We require a mechanism to reproduce similar results with later runs of the same code and  primitives for preventing compiler optimization for particular pieces of code as needed.  For visualization, we desire a primarily web-based tool that allows admins to configure default views of the data but also allows users  to define custom dashboards. Given we are comparing performance results against a historical record, the backend database will need strong support for time-series-oriented data.

For our performance monitoring primitives, we considered the Google benchmark library \cite{gbench}, Celero \cite{celero}, Nonius \cite{nonius} and hayai \cite{hayai} frameworks. Nonius is a small open-source framework for benchmarking small parts of C++ code, inspired by Criterion (a similar Haskell-based tool). Celero, Nonius, hayai have the smaller user community with no clear perspective of continuous support of products, comparing to Google benchmark broader user community.

Googlebench is C++ library, inspired by googletest \cite{gtest} and the xUnit \cite{xunit} architecture.  This library is lightweight and supports different type of benchmarks: both value- and type-parameterized. Google benchmark also provides various options for running the benchmarks including multithreading, and custom report generation.

A key ingredient for benchmarking infrastructure is modern visualization tool that requires minimal maintenance efforts. Visualizing data helps to monitor the performance of ROOT, detect patterns, and take action when identifying anomalous behavior. There are a remarkable number of different data visualization tools used for operations with collected data logs and measurements; we focus on two of the currently most popular and competitive products: Kibana \cite{kibana} and Grafana \cite{grafana}.

Kibana is an open source, browser-based analytics and search dashboard for ElasticSearch \cite{elasticsearch}. Kibana allows for deep explorations of the data, creation and sharing of visualizations, and creation of dashboards; we found it to be both flexible and powerful. Grafana is a general purpose dashboard and graph composer. It is focused on providing rich ways to visualize time series metrics, mainly through graphs but does provide addition tools. It supports multiple databases backends via a plugin mechanism, including InfluxDB \cite{influxdb}.

We decided to utilize the Google benchmarking library because it gives improved control over hotspot functions in ROOT and related libraries. This provides detailed data measurements, including memory usage and CPU instruction counters. Additionally, the framework manages traditional benchmarking pitfalls via repeating unstable benchmarks and providing a stable performance environment over time. Finally, it has a license compatible with ROOT and a large user including a major technology company (Google).

For visualization, we selected Grafana because of its offered feature set and because the host lab for ROOTBench, CERN, provides an excellent on-demand service for Grafana. In turn, this led us to select InfluxDB as our time series database, given its native support in Grafana. The setup allows us to store a sequence of measured data points, providing data for the same performance measurements over time. Each of our performance measurement recorded in InfluxDB is timestamped and defined by a name and a set of labeled dimensions (or “tags”).



\section{Information Flow in ROOTBench} \label{sec:info-flow}
\begin{figure}[!h]
  \centering
  \begin{tikzpicture}[outer sep=0.05cm, node distance=0.8cm, scale=0.7, transform shape]
        
    \node[model, fill=my_green, name=nt] (nt) {Nightly trigger};
    \node[model, fill=my_green, name=pr, right=1cm of nt] (pr) {PR trigger};
    \node[model, fill=my_orange, name=jj, below=1cm of nt] (jj) {Jenkins job};
    \node[model, fill=my_purple, name=git, below=1cm of jj] (git) {rootbench.git/root.git};
    \node[model, fill=my_purple, name=gitpr, right=1cm of git] (gitpr) {Opening PR vs rootbench.git};
    \node[model, fill=my_yellow, name=influxdb, below=1cm of git] (influxdb) {InfluxDB};
    \node[model, fill=my_lightblue, name=grafana, right=1cm of influxdb] (grafana) {Grafana};
    \node[model, fill=my_lightblue, name=def_vis, below=1cm of grafana] (def_vis) {Default Visualization};
     \node[model, fill=my_lightblue, name=custom_vis, right=1cm of def_vis] (custom_vis) {Custom Visualization};

    \draw[line, ->] (nt.south) -- (jj);
    \draw[dashed, ->] (pr.south) -- (jj);
    \draw[line, ->] (jj.south) -- (git);
    \draw[line, ->] (gitpr) -- (git);
    \draw[line, ->] (git.south) -- (influxdb);
    \draw[line, ->] (influxdb.east) -- (grafana);
    \draw[line, ->] (grafana.south) -- (def_vis);
    \draw[dashed, ->] (grafana.south) -- (custom_vis);

  \end{tikzpicture}
  \caption{ROOTBench data flow.}
  \label{fig:InformationFlow}
\end{figure}
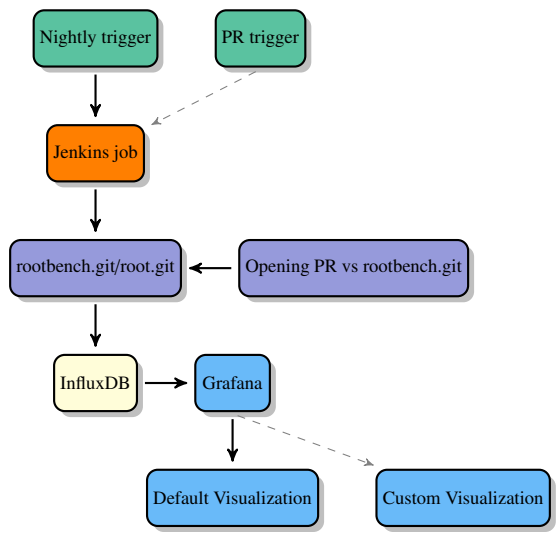

Utilizing Google benchmark, Grafana, and InfluxDB, we have built ROOTBench \cite{rootbench} to help understand ROOT code performance and monitor the efficiency of the code for several processor architectures; in this section, we explain how data measurements are triggered and eventually are recorded and visualized.  This is outlined in Figure \ref{fig:InformationFlow}.

The data measurements in the system starts from a set of user-defined benchmarks written using the Google benchmark framework and compiled to a standalone executables. All benchmarks are compiled by a nightly job using the Jenkins framework. The Jenkins job will compile both the ROOT framework and the ROOTbench software repository against the ROOT build. After the builds have completed successfully, the set of benchmarks in sequential mode on a dedicated machine (done to reduce the performance fluctuations). After each benchmark, the performance data measurements are uploaded into an InfluxDB database.

The visualization system is only coupled to the continuous measurements via the database as shown in Figure \ref{fig:InformationFlow}. Grafana reads the database and visualizes the obtained data, making it available to the developer community via the Grafana web interfaces. Note that, other than the ROOT-specific benchmarks themselves, the approach taken by this framework is relatively generic and can be reused for other projects.



\section {Usage of ROOTbench} \label{sec:usage}

We believe writing micro-benchmarks (focusing on a single function or small piece of functionality) instead of monolithic macro benchmarks (covering a wide swatch of ROOT's feature set) is more relevant for monitoring performance of software hotspots. We have found micro benchmarks are hard to write and easy to debug, whereas macro benchmarks are easy to write and hard to debug. Given changes to macro benchmark results can be caused by a number of software hotspots, we find they need to be coupled with extra profiling and debugging tools such as Intel VTune \cite{vtune}. Thus, tracking performance regressions with macro benchmarks can require extra human investigation time. Finally, macro benchmarks can hide performance regressions given their overall execution time; the size of the regression may be larger than the jitter in the benchmark performance. A micro benchmark, by contrast, provides a routine which checks performance only for some concise, targeted functionality. The call stack and code coverage of the benchmark should be as small as possible. The concept of a micro benchmark is thus similar to that of unit test. This kind of benchmarks is very sensitive and can efficiently amplify and outline regressions. Assuming measurement jitter is relatively bounded, finding regressions becomes easier.

It can be challenging for a micro-benchmarking infrastructure to provide stable measurements over time, especially as this may require the same physical machines be consistently used for an extended time (to provide the broadest set of historical data).


\subsection{Examples}

We want to introduce to the reader some sample microbenchmarks for the ROOTBench system; for checking scalability, we present two examples, concerning vectorization and threading.

\begin{listing}[h]
    \noindent
    \begin{minipage}[h]{.7\textwidth}
   \begin{cppcode*}{}
  #include <benchmark/benchmark.h>
  #include "Math/GenVector/Transform3D.h"
  #include "Math/Types.h"

  template <typename T>
  static void BM_Plane3D_Hessian(benchmark::State &state) {
     const double a(p0(ggen)), b(p1(ggen)), c(p2(ggen)), d(p3(ggen));
     for (auto _ : state) {
        Plane<T> sc_plane(a, b, c, d);
        sc_plane.HesseDistance();
     }
   }
   BENCHMARK_TEMPLATE(BM_Plane3D_Hessian, double)->Unit(benchmark::kMicrosecond);
   BENCHMARK_TEMPLATE(BM_Plane3D_Hessian, ROOT::Double_v)->Unit(benchmark::kMicrosecond);
   BENCHMARK_MAIN();
   \end{cppcode*}
   \end{minipage}
   \caption{Monitoring vectorization scalability of Plane3D::Hessian}
   \label{vec_bench}
\end{listing}

Listing~\ref{vec_bench} starts with the inclusion of the benchmarking primitives and the checked entities. Then we define a function which will monitor the \textit{GenVector}'s \textit{Plane3D::HesseDistance}. After the initialization section, we use a loop which is controlled by the framework. Its purpose is to stabilize the results. Next two macro expansions \textit{BENCHMARK\_TEMPLATE} are used for registering our \textit{BM\_Plane3D\_Hessian} as a monitoring function. We register two instances of the function, one taking standard type \textit{double} and a second one taking a vectorization-enabling type \textit{ROOT::Double\_v}. This allows us to compare the level of vectorization against the baseline. We finish by defining the benchmark executable main function via \textit{BENCHMARK\_MAIN}.

Listing~\ref{thread_bench} defines a multi-threaded benchmark. Line 7, 8, 20 and 21 show the initialization and destruction sections. In the loop, we monitor a small helper function which uses the \textit{TBufferMerger} API to fill a \textit{TTree} with random data. Line 11 makes the benchmark configurable by allowing to set the flush level by specifying a \textit{Range} configuration.

Function is registered as a benchmark with parameters of flush rates between 1 and 32 MB with a step of 8. We configure the benchmark to use real time which is more critical in multithreaded environments. We set a thread range in the interval between 1 and twice the number of underlying hardware threads to measure hyperthreading effects. The visualization of a similar  example function can be seen in Figure~\ref{tbuf}.

\begin{listing}[h]
   \noindent
   \begin{minipage}[h]{.7\textwidth}
   \begin{cppcode*}{}
   // ...
   BufferMerger *Merger = nullptr;
   static void BM_TBufferFile_FillTree(benchmark::State &state) {
      ROOT::EnableThreadSafety();
      using namespace ROOT::Experimental;
      // Setup code here.
      if (state.thread_index == 0)
         Merger = new TBufferMerger(std::unique_ptr<TMemFile>(new TMemFile("v_file.root",
                                                                           "RECREATE")));
      long size;
      int flush = state.range(0);
      for (auto _ : state) {
         FillTreeWithRandomData(*Merger, flush);
         size = Merger->GetFile()->GetSize();
      }
      std::stringstream ss;
      ss << size;
      state.SetLabel(ss.str());
      // Teardown code here.
      if (state.thread_index == 0)
         delete Merger;
   }
   BENCHMARK(BM_TBufferFile_FillTree)->Range(1, 32)->UseRealTime()
                                     ->ThreadRange(1, GetNumberHardwareThreads() * 2);
   \end{cppcode*}
   \end{minipage}
   \caption{Monitoring threading scalability of ROOT's TBufferMerger}
   \label{thread_bench}
\end{listing}

\section{Results} \label{sec:results}

Each Grafana panel is tied to a specific data source that belongs to a particular Grafana organization.  The dashboard is where it all comes together: data via queries, panels, and its settings. Dashboards can be considered as a set of one or more panels organized into rows. A user can control the period of interest by selecting it in the dashboard time picker. This gives the possibility to visualize performance statistics from hours to months. Dashboards can be dynamic, which allows a single template to be applied to multiple sets of data with similar properties. More complex dashboards can be used to correlate the time series data in the panel with other external factors, such as an operating system update or other performance-significant changes significant (via annotations). Dashboards (or a specific panel) can be shared via unique URLs, or via the snapshot, functionality to encode all the data currently being viewed into a file.

\begin{figure}[h]
\centering
\includegraphics[width=\linewidth]{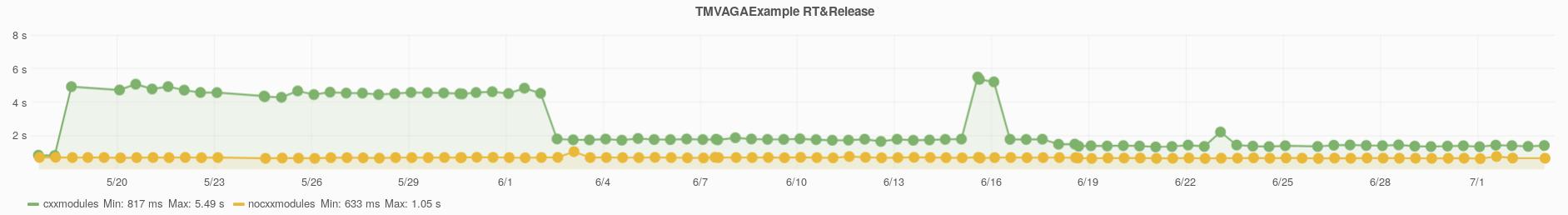}
\includegraphics[width=\linewidth]{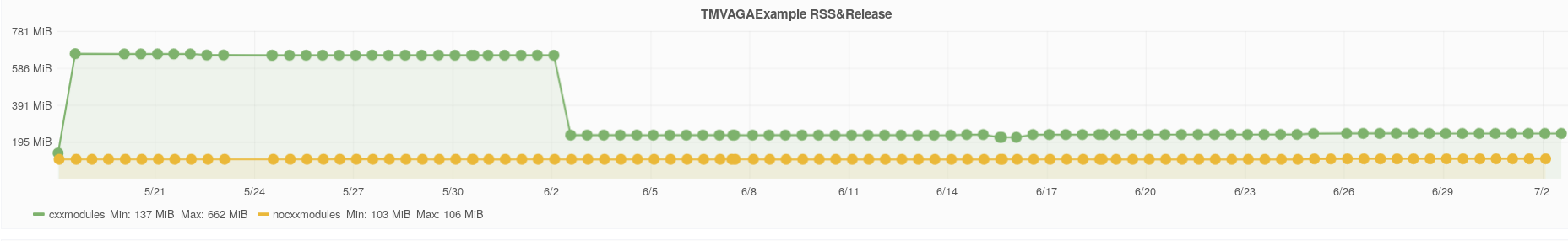}
\includegraphics[width=\linewidth]{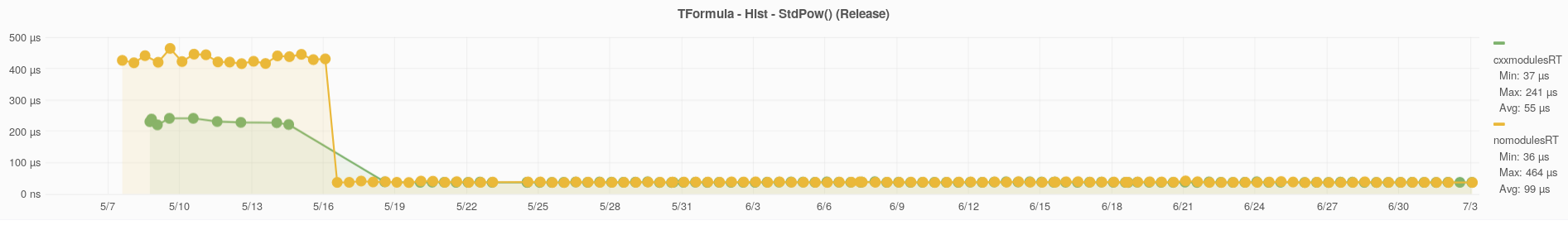}
\caption{Performance monitoring plots for ROOT's runtime\_cxxmodules and pch features.}
\label{interp}
\end{figure}

\begin{figure}[h]
\centering
\includegraphics[width=\linewidth]{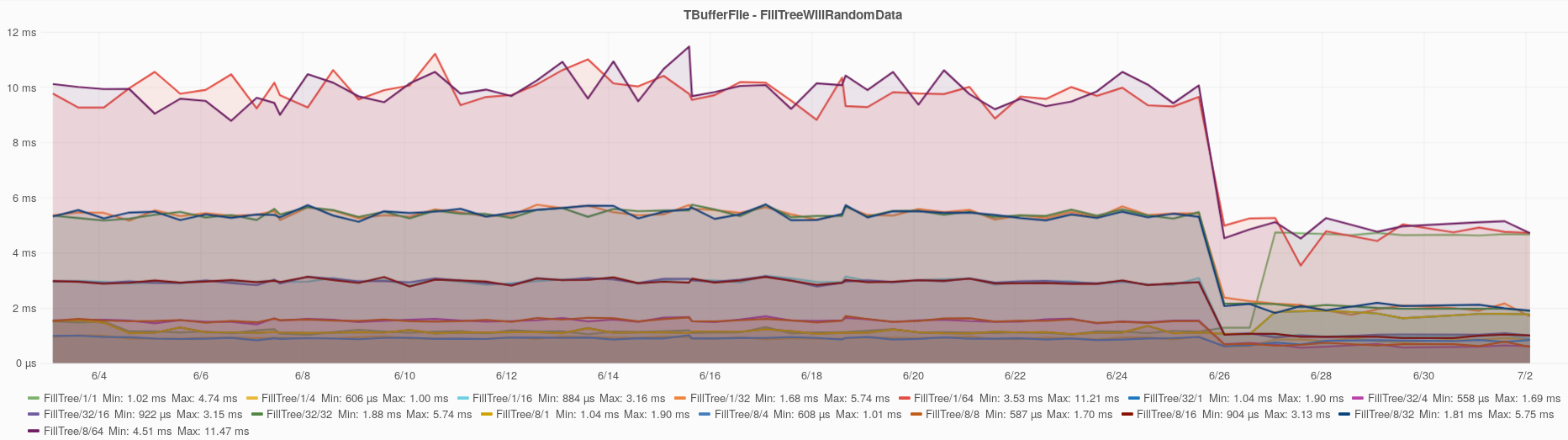}
\caption{Threading Improvements in TBufferMerger.}
\label{tbuf}
\end{figure}

\begin{figure}[h]
\centering
\includegraphics[width=\linewidth]{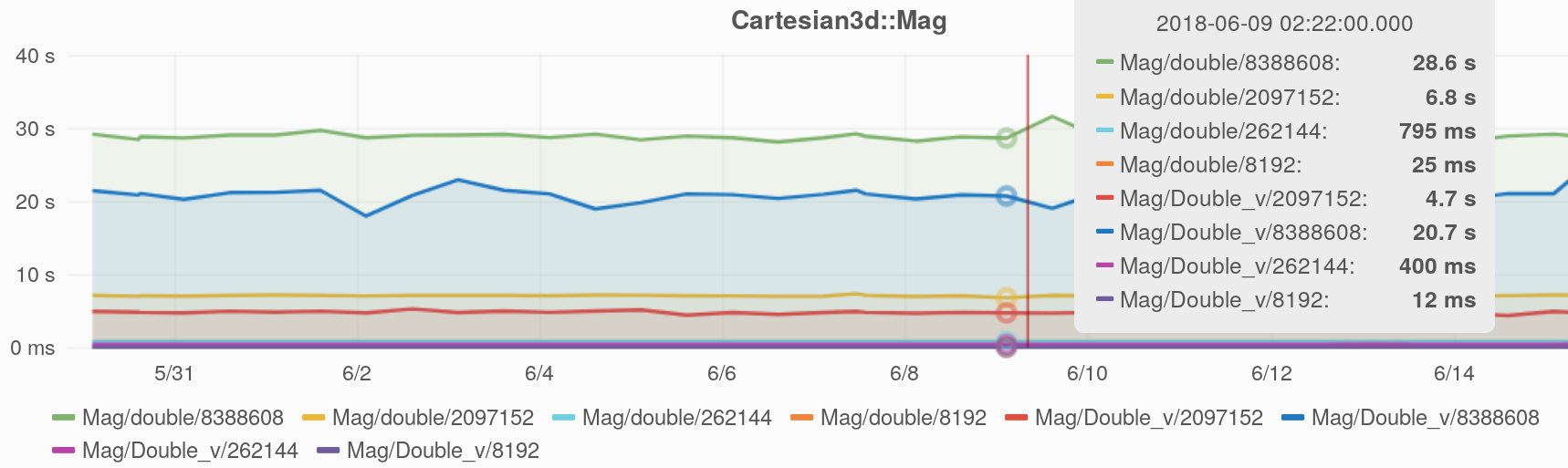}
\caption{Vectorization Scalability of GenVector::Mag() function.}
\label{fig:genvec}
\end{figure}

\begin{figure}[h]
\centering
\includegraphics[width=\linewidth]{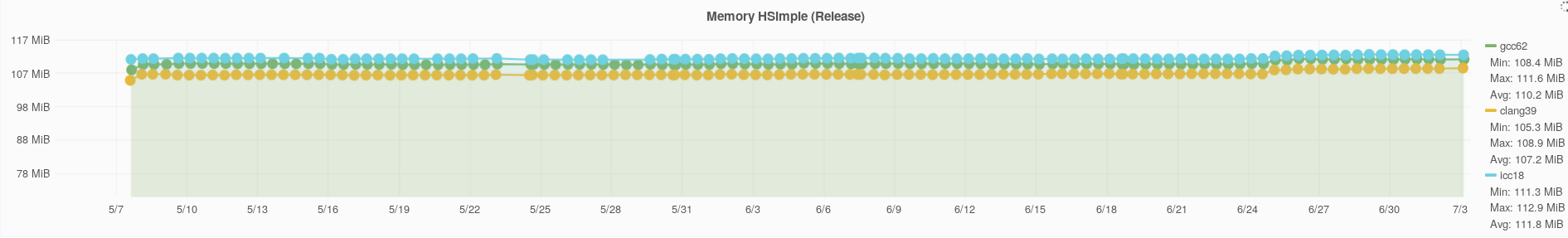}
\caption{Memory Footprint of clang-, icc- and gcc-compiled ROOT Running HSimple.}
\label{fig:compilers}
\end{figure}

Currently we define Real Time (RT), CPU Time and  memory usage footprint measurements. Figures \ref{interp}, \ref{tbuf}, \ref{fig:genvec}, \ref{fig:compilers} show several real-world examples from Grafana instance at https://root-bench.cern.ch.

Figure \ref{interp} outlines performance improvements in the Interpreter benchmarks. In green, we denote the performance of the ROOT C++ modules feature \cite{modules} and in yellow is the default ROOT build. In this case, the improvement in C++ modules is due to multiple memory optimizations and implementation of more efficient symbol resolution at runtime.

Figure~\ref{tbuf} outlines recent performance optimizations in the \textit{TBufferMerger} class. A callback functionality was removed to avoid thread oversubscription. The improvements made ROOT more thread-safe, and now users can decide between use threads and tasks with TBufferMerger. Figure~\ref{fig:genvec} demonstrates a comparison of vectorized and scalar code in the \textit{Mag} function of ROOT's \textit{GenVector} library. Figure~\ref{fig:compilers} shows the memory footprint of the \textit{HSimple} benchmark between ROOT builds from different compilers.

\section {Conclusion} \label{sec:conclusions}

The ROOTBench project is still in early phases, but has proved shown to be useful. The tool has been often used to identify performance degradation. Performance-sensitive code can be monitored at the cost of a GitHub ``pull request" against the ROOTBench GitHub repository. ROOTBench is currently a part of the ROOT release checklist, a necessary verification step to ensure there are no performance regressions from the current release. 

Along with the list of existing features, the ROOTBench team has a long list of items remaining:
\begin{itemize}
  \item enable the current infrastructure to work on every ROOT pull request (likely, this requires additional hardware resources).
  \item addition of more benchmarks.
  \item add performance node monitoring to monitor regressions in the host hardware.
  \item add version control to web dashboards.
  \item addition of custom metrics for benchmarks, such as precise memory footprints from cling AST statistics.
  \item addition of alert mechanisms (in cases where benchmark values go beyond a specified threshold).
    \item add mechanisms to run on older commits 
    \item separating the coverage measurements based on additive coverage calculation.
    \item addition of annotations -- areas from the Grafana plots which can be marked false positives by users
\end{itemize}

Performance of large-scale systems is fragile and can vary on the different systems. It is vital for  projects to offer a set of tools and benchmarks allowing coders to reason about performance. We believe ROOTBench is a step towards recognizing and solving the problem ensuring better sustainability of HEP Software.

{\small This work has been supported by an Intel Parallel Computing Center grant and by U.S. National Science Foundation grant OAC-1450377, OAC-1450323, and PHY-1624356.
The  authors would like to  thank CERN OpenLab for providing dedicated machines for continuous performance monitoring.}


\begin{thebibliography}{plain}

\bibitem{root}
R. Brun, F. Rademakers, \textit{ROOT - An Object Oriented Data Analysis Framework}, Nucl. Inst. \& Meth. in Phys. Res. A  \textbf{389} (Proceedings AIHENP'96 Workshop,1997).

\bibitem{gbench}
GitHub. 2018. GitHub - google/benchmark: A microbenchmark support library. Available at: https://github.com/google/benchmark/. [Accessed 29 November 2018].

\bibitem{celero}
GitHub. 2018. GitHub - DigitalInBlue/Celero: C++ Benchmark Authoring Library/Framework. Available at: https://github.com/DigitalInBlue/Celero. [Accessed 29 November 2018].

 \bibitem{nonius}
 GitHub. 2018. GitHub - libnonius/nonius: A C++ micro-benchmarking framework. Available at: https://github.com/libnonius/nonius. [Accessed 29 November 2018].
 
 \bibitem{hayai}
 GitHub. 2018. GitHub - nickbruun/hayai: C++ benchmarking framework. Available at: https://github.com/nickbruun/hayai. [Accessed 29 November 2018].
 
 \bibitem{gtest}
GitHub. 2018. GitHub - abseil/googletest: Google Test. Available at: https://github.com/google/googletest.git. [Accessed 29 November 2018].
 
 \bibitem{xunit}
 Xunit.github.io. (2019). xUnit.net. [online] Available at: https://xunit.github.io/ [Accessed 6 Feb. 2019].
 
 \bibitem{kibana}
 Elastic. 2018. Kibana: Explore, Visualize, Discover Data | Elastic. Available at: https://www.elastic.co/products/kibana/. [Accessed 29 November 2018].
 
\bibitem{grafana}
Grafana Labs. 2018. Grafana - The open platform for analytics and monitoring. Available at: https://grafana.com/. [Accessed 29 November 2018].
 
 \bibitem{elasticsearch}
 Gormley, Clinton, and Zachary Tong. Elasticsearch: The Definitive Guide: A Distributed Real-Time Search and Analytics Engine. " O'Reilly Media, Inc.," 2015.
 
\bibitem{influxdb}
InfluxData. 2018. InfluxDB 1.7 documentation | InfluxData Documentation. Available at: https://docs.influxdata.com/influxdb. [Accessed 29 November 2018].

\bibitem{rootbench}
GitHub. 2018. GitHub - root-project/rootbench: Collection of benchmarks and performance monitoring applications. Available at: https://github.com/root-project/rootbench/. [Accessed 29 November 2018].

\bibitem{vtune}
Intel® VTune™ Amplifier 2019 User Guide. Available at: https://software.intel.com/en-us/vtune-amplifier-help [Accessed 29 Nov. 2018].

\bibitem{modules}
Y. Takahashi, V.Vassilev, O.Shadura, R.Isemann. \textit{Optimizing Frameworks Performance Using C++ Modules Aware ROOT.} CoRR abs/1812.03992 (2018)

\end{thebibliography}
\end{document}